\documentclass[prd,aps,showpacs,preprintnumbers,amssymb]{revtex4}
\usepackage{axodraw}
\usepackage{color}
\usepackage{epsf}

\def\e3p{$\eta \rightarrow 3 \pi$}

\begin{document}
\title{%
\hfill{\normalsize\vbox{%
\hbox{}
 }}\\
{ Why are there three generations of fermions in the standard model?}}

\author{Amir H. Fariborz
$^{\it \bf a}$~\footnote[1]{Email:
 fariboa@sunyit.edu}}
\author{Renata Jora
$^{\it \bf b}$~\footnote[2]{Email:
 rjora@theory.nipne.ro}}

\author{Joseph Schechter
 $^{\it \bf c}$~\footnote[4]{Email:
 schechte@phy.syr.edu}}

\affiliation{$^{\bf \it a}$ Department of Mathematics/Physics, State University of New York, Polytechnic Institute,  Utica, NY 13504-3050, USA}
\affiliation{$^{\bf \it b}$ National Institute of Physics and Nuclear Engineering PO Box MG-6, Bucharest-Magurele, Romania}

\affiliation{$^ {\bf \it c}$ Department of Physics,
 Syracuse University, Syracuse, NY 13244-1130, USA}

\date{\today}

\begin{abstract}
We show that by decomposing the gauge fields in fermion degrees of freedom and by saturating the remaining degrees of freedom as dynamical fields in the Lagrangian one might explain the
proliferation of fermion states in the standard model Lagrangian. Thus the mere presence of the gauge symmetry $U(1)_Y \times SU(2)_L \times SU(3)_c$ is essential.
\end{abstract}
\pacs{12.15.Ff,12.90.+b}
\maketitle

The standard model of elementary particles \cite{Glashow}-\cite{Veltman} contains three generations of fermions each consisting of 15 two component fermion fields. For example the first generation contains $\nu_{L}$, $e_L$, $e_R$, $u_L$, $u_R$ , $d_L$, $d_R$ where one should count also for the three color states for each quark. Solutions have been proposed to explain the presence of three generations, among which one of the most significant is the extension of the standard model gauge group to groups like $SU(5)$ or $SO(10)$  in an attempt to unify all of the fundamental interactions. However up to today there is no widely accepted and convincing answer to this problem.
In this work we shall give a simple, concise and self consistent response to why there are three generations of fermions using only the limited context of the standard model without the need of introducing additional particles and interactions.

We shall analyze the $U(1)_{em}$ part of the standard model after spontaneous symmetry breaking but the same arguments apply as well to $U(1)_Y$ with the amend that one should consider separately the left handed and right handed states. The Lagrangian of interest reads:
\begin{eqnarray}
{\cal L}=-\frac{1}{4}F^{\mu\nu}F_{\mu\nu}+e \sum_f Q(f)A_{\mu}\bar{f}\gamma^{\mu}f+...,
\label{lagr33245}
\end{eqnarray}
where the sum goes over all charged fermions and $Q(f)$ is the respective charge.
We choose the Feynman gauge ($\xi=1$) and write the equation of motion for the abelian gauge field:
\begin{eqnarray}
-\partial^{\mu}\partial_{\mu}A^{\nu}-\sum_fe Q(f)\bar{f}\gamma^{\nu}f=0.
\label{new6}
\end{eqnarray}

We neglect the Higgs doublet in what follows as they are not essential for our arguments. We shall rewrite without any loss of generality the electromagnetic field as:
\begin{eqnarray}
A^{\nu}=k^2[\bar{\Psi}\gamma^{\nu}\chi+\bar{\chi}\gamma^{\nu}\Psi],
\label{write54678}
\end{eqnarray}
where k is a constant parameter that has dimension of inverse mass.
First we need to justify the choice that we made in Eq. (\ref{write54678}). The most general way in which one can write a neutral vector field in terms of two Dirac spinors is:
\begin{eqnarray}
A^{\nu}=k^2[a_1\bar{\xi}_1\gamma^{\nu}\xi_1+a_2\bar{\xi}_2\gamma^{\nu}\xi_2],
\label{rec454}
\end{eqnarray}
where $a_1$ and $a_2$ are real coefficients with mass dimension zero. Then one can observe that Eq.(\ref{write54678}) upon a change of variable is a particular case of Eq. (\ref{rec454}) with $\xi_1=\Psi+\chi$ and $\xi_2=\Psi-\chi$. The specific choice we make does not influence in any way the generality of the arguments we propose in this work.  Second we need to explain why we consider that the minimum choice for $A^{\nu}$ contains at least two spinors instead of one.  This stems from the electroweak group $U(1)_Y\times SU(2)_L$ that contains four generators. The gauge fields contain  eight real degrees of freedom from which one can extract after spontaneous symmetry breaking two neutral states and two charged ones. It is then obvious that these states must be expressed in terms of at least two fermions. Since $U(1)_{em}$ is obtained as a linear combination of two of the initial gauge fields one expects that generally it contains in its structure both fermion states. Thus our choice is completely justified in the standard model context although it is not the most general one.

Note that at this stage one should not identify $\Psi$ and $\chi$ with any of the standard model fermions. Also  Eq. (\ref{write54678})  does not necessarily imply that the field $A^{\nu}$ is composite but is only a way of rewriting the gauge field in terms of new variables. We then apply the operator $\partial_{\mu}\partial^{\mu}$ to the field $A^{\nu}$ to get:
\begin{eqnarray}
&&\frac{1}{k^2}\partial^{\mu}\partial_{\mu}A^{\nu}=
\nonumber\\
&&\partial^{\mu}\partial_{\mu}\bar{\Psi}\gamma^{\nu}\chi+\bar{\chi}\gamma^{\nu}\partial_{\mu}\partial^{\mu}\Psi+
\nonumber\\
&&\bar{\Psi}\gamma^{\nu}\partial^{\mu}\partial_{\mu}\chi+\partial^{\mu}\partial_{\mu}\bar{\chi}\gamma^{\nu}\Psi+
\nonumber\\
&&\partial^{\mu}\bar{\Psi}\gamma^{\nu}\partial_{\mu}\chi+\partial^{\mu}\bar{\chi}\gamma^{\nu}\partial_{\mu}\Psi.
\label{expr4567}
\end{eqnarray}
We denote,
\begin{eqnarray}
&&k^2\partial^{\mu}\partial_{\mu}\Psi=\Psi_1
\nonumber\\
&&k^2\partial^{\mu}\partial_{\mu}\chi=\chi_1
\nonumber\\
&&k^2\partial^{\mu}\bar{\Psi}\gamma^{\nu}\partial_{\mu}\chi=\bar{\Psi_2}\gamma^{\nu}\chi_2.
\label{not5677}
\end{eqnarray}
We are allowed to write the last line in Eq. (\ref{not5677}) as we did abecause we know that the corresponding term behaves as a Lorentz vector.

Then Eq. (\ref{expr4567}) can be rewritten as:
\begin{eqnarray}
&&2\partial^{\mu}\partial_{\mu}A^{\nu}=
\nonumber\\
&&(\bar{\Psi}_1+\bar{\chi})\gamma^{\nu}(\Psi_1+\chi)-(\bar{\Psi}_1-\bar{\chi})\gamma^{\nu}(\Psi_1-\chi)+
\nonumber\\
&&(\bar{\Psi}+\bar{\chi}_1)\gamma^{\nu}(\Psi+\chi_1)-(\bar{\Psi}-\bar{\chi}_1)\gamma^{\nu}(\Psi-\chi_1)+
\nonumber\\
&&(\bar{\Psi}_2+\bar{\chi}_2)\gamma^{\nu}(\Psi_2+\chi_2)-(\bar{\Psi}_2-\bar{\chi}_2)\gamma^{\nu}(\Psi_2-\chi_2)
\label{rez32456}
\end{eqnarray}
Now if one compares Eq. (\ref{new6}) with Eq. (\ref{rez32456}) one notes that one can account at least for six  charged fermions, three with positive charges and three with negative ones. We shall consider these as being $e$, $\mu$, $\tau$, $u$, $c$ and $t$.  Of course Eqs. (\ref{new6}) and (\ref{rez32456}) may contain un unlimitted number of fermions terms. However we claim that this is the minimum number that must exist. The down quark states are then due to the presence of the $SU(2)_L$ group and its corresponding quantum numbers. The same arguments may be applied also to any vector group and in the case of $SU(3)_c$ indicate that one needs at least 6 flavors of fermions charged under color.

Now we should explain rigourously why one must have  in Eq. (\ref{rez32456}) at least six independent fermion terms corresponding to at least six species of charged fermions. For that we consider Eq. (\ref{write54678}) and note that it contains four equation corresponding to the four space time components. On the other hand we have two Dirac spinors each having four complex components which in total would correspond to sixteen real components.  This means that after solving the equations we are left with twelve independent degrees of freedom or six complex ones. Next we need to show that six complex degrees of freedom correspond to at least six terms in Eqs. (\ref{new6}) and (\ref{rez32456}) and thus to six species of fermions. For that we consider the equation of motion for an arbitrary fermion $f$ with the charge $Q(f)=-1$:
\begin{eqnarray}
\gamma^{\mu}\partial_{\mu}f+i e A_{\mu}\gamma^{\mu}f=0
\label{ferm789}
\end{eqnarray}
This equation contains four complex components or eight real ones which completely determines the spinor. The number of constraints on the field f which are independent on $A_{\mu}$ reduces to four. Thus the constrained fermion field contains four real independent degrees of freedom. However the phase space of a fermion field (\cite{Tong}) is parameterized by $f $ and $f^{\dagger}$ so for the constrained fermion contains four real degrees of freedom. This means that the corresponding number of degrees of freedom for the constrained fermion is half of it which is two real degrees of freedom. On the right hand side of Eq. (\ref{rez32456}) we have twelve independent degrees of freedom and this  should correspond  as indeed the number of terms suggest to six different fermion species.

A few comments need to be made in conclusion. Eq. (\ref{write54678})  represents a general reparametrization of the gauge field in terms of two Dirac fermions. Nevertheless there is an infinite number of ways in which the field $A_{\mu}$ can be reexpressed. However we select from these the most general decomposition in term of two Dirac fields.  Moreover even if the right hand side contains more independent degrees of freedom this do not need to actually be saturated in the Lagrangian.  We made here an assumptions that we shall call it the principle of minimum decomposition which is: If we rewrite a gauge field in the most general way in terms of fermion components or other new field variables then  the remaining independent degrees of freedom  should be present in the Lagrangian as dynamical fields.  This principle solely together with the gauge symmetry $U(1)_Y\times SU(2)_L \times SU(3)_c$ can explain the number of generations in the standard model Lagrangian.

\section*{Acknowledgments} \vskip -.5cm
The work of R. J. was supported by a grant of the Ministry of National Education, CNCS-UEFISCDI, project number PN-II-ID-PCE-2012-4-0078.


\begin{thebibliography}{15}
\bibitem{Glashow} S. L. Glashow, Nuxl. Phys. {\bf 22} (4): 579-588 (1961).
\bibitem{Weinberg} S. Weinberg, Phys. Rev. Lett.  {\bf 19} (21):1264-1266 (1967).
\bibitem{Brout} F. Englert and R. Brout, Phys. Rev. Lett. {\bf 13} (16):321-323 (1964).
\bibitem{Higgs} P. W. Higgs, Phys. Rev. Lett.  {\bf 13}  (16): 508-509 (1964).
\bibitem{Salam} A. Salam, Elementary Particle Theory, Nobel Symposium No. 8, N. Svartholm (eds) (Almqvist and Wiksells, Stockholm 1968), p. 137.
\bibitem{Guralnik} G. S. Guralnik, C. R. Hagen and T. W. B. Kibble, Phys. Rev. Lett. {\bf 13}, 585 (1964).
\bibitem{Hooft} G. 't Hooft, Nucl. Phys. B {\bf 3}, 167 (1971).
\bibitem{Veltman} G. 't Hooft and M. J. G. Veltman, Nucl. Phys. B {\bf 44}, 189 (1972).
\bibitem{Tong} D Tong, Lectures on Quantum Field Theory, Lectures notes copyright@2006 David Tong.
\end{thebibliography}
\end{document}